\documentclass[preprint,superscriptaddress,prd,showpacs]{revtex4}
\usepackage{amssymb,amsmath,graphicx,amscd,xcolor}

\def\E{\mathbf{E}}
\def\B{\mathbf{B}}
\def\D{\mathbf{D}}
\def\H{\mathbf{H}}
\def\P{\mathbf{P}}
\def\x{\mathbf{x}}
\def\k{\mathbf{k}}

\begin{document}

\title{An Analog Model for Light Propagation in Semiclassical Gravity}

\author{C. H. G. Bessa}
\email{carlos@cosmos.phy.tufts.edu}
\affiliation{Institute of Cosmology, Department of Physics and Astronomy,
Tufts University, Medford, Massachusetts 02155, USA}

\author{V. A. De Lorenci}
\email{delorenci@unifei.edu.br}
\affiliation{Instituto de Ci\^encias Exatas \\ Universidade Federal de Itajub\'a, Itajub\'a, MG 37500-903, Brazil}
\affiliation{Institute of Cosmology, Department of Physics and Astronomy,
Tufts University, Medford, Massachusetts 02155, USA}

\author{L. H. Ford}
\email{ford@cosmos.phy.tufts.edu}
\affiliation{Institute of Cosmology, Department of Physics and Astronomy,
Tufts University, Medford, Massachusetts 02155, USA}

\begin{abstract}
We treat a model based upon nonlinear optics for the semiclassical gravitational effects of quantum fields upon 
light propagation. Our model uses a nonlinear material with a nonzero third order polarizability. Here
a probe light pulse  satisfies a wave equation containing the expectation value of the squared electric field.
This expectation value depends upon the presence of lower frequency quanta, the background field, and modifies the effective
index of refraction, and hence the speed of the probe pulse. If the mean squared electric field is positive, then the pulse
is slowed, which is analogous to the gravitational effects of ordinary matter. Such matter satisfies the null energy
condition and produce gravitational lensing and time delay. If the mean squared field is negative, then the pulse
has a higher speed than in the absence of the background field. This is analogous to the gravitational effects of
exotic matter, such as stress tensor expectation values with locally negative energy densities, which lead to
repulsive gravitational effects, such as defocussing and time advance. We give some estimates of the magnitude
of the effects in our model, and find that they may be large enough to be observable. We also briefly discuss the
possibility that the mean squared electric field could be produced by the Casimir vacuum near a reflecting boundary. 
 \end{abstract}
		
\maketitle
\baselineskip=14pt	

\section{Introduction}
\label{sec:intro}

Semiclassical gravity, in which the renormalized expectation value of a matter stress tensor operator is the source
of a classical gravitational field, reproduces all of the phenomena of classical general relativity. 
The effects of classical gravitational fields upon the propagation of light rays include gravitational lensing and
the Shapiro time delay~\cite{Shapiro}. More precisely, these are the effects of gravitational fields produced by matter
which satisfies classical energy conditions, such as the null energy condition, $T_{\mu\nu} k^\mu k^\nu \geq 0$
for all null vectors $k^\mu$. However, semiclassical gravity also describes the gravitational effects of matter
which violates this condition, as may occur when negative energy density is present.
If the classical conditions are violated, then defocussing and time advance effects are
possible. The simplest example would be a negative mass Schwarzschild geometry, which would act as a diverging
lens and exhibit a Shapiro time advance. The connection between exotic matter and superluminal propagation has
been discussed by several authors~\cite{PSW,Olum,VBL}, who show that violation of the null energy condition is
required to have a time advance, a light ray traveling faster than it would in flat spacetime. Quantum field theory
does allow local negative energy densities, as in the Casimir effect or in squeezed states, but these effects are 
very restricted in either space or time. In the case of the Casimir effect with parallel perfect plane mirrors at a separation 
of $\ell$, there is a uniform static negative energy density between the mirrors of $-\hbar c /(720 \pi^2 \, \ell^4)$. In the 
case of more realistic mirrors, it is possible still to have a reduced negative energy density~\cite{SF02,SF05}. In the case
of nonclassical quantum states, such as squeezed vacua, the negative energy is limited in its temporal duration
by quantum inequalities~\cite{ford78,ford91,fr95,fr97,flanagan97,FE98,pfenning01,fh05}. 
These inequalities give an inverse
relation between the magnitude and duration of negative energy in the frame of any inertial observer, and 
greatly limit the gravitational effects of negative energy density.

However, the quantum inequalities do not mean that negative energy density, or related subvacuum effects, are 
unobservable. Subvacuum effects can produce changes in the magnetization of a
spin system~\cite{fgo92} and the rate of atomic decays~\cite{FR11}. In the present paper, we will discuss another
possibility, the effects of negative mean squared electric field upon the propagation of light in a nonlinear material.
The basic idea is that in a material with nonzero third order susceptibility, the speed of propagation of a probe pulse
can depend upon the squared electric field, which can be produced by a background field. This effect is an analog model
for the effects of gravity on light propagation. If the mean squared electric field is positive, $\langle E^2 \rangle > 0$,
then there is an increase in the index of refraction and hence a decrease in the speed of the probe pulse. This
is analogous to the effect of the gravitational field of normal matter on light rays. However,  $\langle E^2 \rangle$
is defined as a difference between values in a given state and in the vacuum, and hence need not be positive. 
 If   $\langle E^2 \rangle < 0$, 
we have a subvacuum effect, and there is an increase in the speed of the probe pulse.  This is analogous to the gravitational
effects of exotic matter. The subvacuum effect can be created by a background field in a squeezed state. A similar
model was used in Ref.~\cite{flms13} to model light cone fluctuations. However, that model used materials with
nonzero second order susceptibility and described fluctuations in propagation speed around an average value 
which was independent of the background field. In the present paper, we are concerned with changes in the mean
speed of light, analogous to effects which occur in the semiclassical theory of gravity.

In Sect.~\ref{sec:max}, some basic information on nonlinear optics will be reviewed, and used to formulate our model,
and to derive expressions for the effective index of refraction in terms of the expectation value of a squared electric
field. These expressions will be evaluated in a multimode squeezed vacuum state in Sect.~\ref{sec:squeeze}. The results
will be used to give some numerical estimates for the possible magnitude of the change in refraction angle due to
subvacuum effects. The change in refractive index due to the Casimir vacuum near a reflecting plate will be discussed
in Sect.~\ref{sec:Casimir}. Our results are summarized and discussed in Sect.~\ref{sec:sum}.

\section{Nonlinear Optics and Light Propagation }
\label{sec:max}

In this section, we review some aspects of electrodynamics in a nonlinear medium and define some  
notation to be used in the following sections.
 In absence of charges and currents Maxwell's equations can be written in SI units as
\begin{eqnarray}
\nabla \cdot \B &=& 0 \,, \quad \nabla\times \E = - \frac{\partial \B}{\partial t}\,, 
\\ 
\nabla\cdot \D &=& 0\,, \quad \nabla\times \H =  \frac{\partial \D}{\partial t} \,,
\label{eq:Maxwell}
\end{eqnarray}
where the induced electric ${\D}$ and magnetic  $\B$ fields are related to the respective intensities $\E$ and $\H$ 
by means of the constitutive relations: $\B = \mu_0 \H $  and  $\D = \epsilon_0\E + \P$. The quantity $\P$ is 
the polarization, which can be expanded in terms of the components of the susceptibility tensors 
\boldmath$\chi$\unboldmath${}^{(a)}$ ($a=1,2,3,...$) as (See, for example, Ref.~\cite{boyd}.)
\begin{equation}
P_i = \epsilon_0 \left(\chi_{ij}^{(1)} E_{j} + \chi_{ijk}^{(2)} E_{j}E_{k} 
+ \chi_{ijkl}^{(3)} E_{j}E_{k}E_{l} + \cdots \right) \,.
\label{eq:pol}
\end{equation}
Here repeated indices are summed over.  We will consider the case in which
$\P$ and $\E$ are parallel. Then the electric field is divergenceless and its evolution is governed by the wave equation,
\begin{equation}
\biggl(\nabla^2 - \frac{1}{c^2}\frac{\partial^2}{\partial t^2}\biggr)\E =  \frac{1}{\epsilon_0 c^2}\frac{\partial^2}{\partial t^2}\P \,.
\label{eq:wave-eq}
\end{equation}
We will further consider only the case where the second order susceptibility vanishes, so $\chi_{ijk}^{(2)} =0$, as
will hold for any material whose crystal lattice possess spatial inversion symmetry.  If we stop the expansion in
Eq.~(\ref{eq:pol}) at third order, then Eq.~(\ref{eq:wave-eq}) contains linear and cubic terms in the electric field.

A key ingredient in our model is to assume that the total electric field may be written as the sum of a background
field ${\bf E}_0$ and a probe field  ${\bf E}_1$,
\begin{equation}
{\mathbf E} = {\mathbf E}_0 + {\mathbf E}_1 \,,
 \end{equation} 
where $|{\bf E}_1| \ll |{\bf E}_0|$ and   $|\nabla \ln(|{\bf E}_1|)| \gg |\nabla \ln(|{\bf E}_0|)|$. That is, the probe field is a
small, but rapidly varying, perturbation of the background field. We can greatly simplify our model by taking both
fields to be linearly polarized in the $z$-direction, so  ${\mathbf E}_0 = E_0(t,x,y){\bf \hat{z}}$ and
${\mathbf E}_1 = E_1(t,x,y){\bf \hat{z}}$.
In this case, the only coefficients of the susceptibility tensors that contribute to the wave propagation are 
$\chi_{zz}^{(1)}$ and $\chi_{zzzz}^{(3)}$, which we will denote as $\chi^{(1)}$ and $\chi^{(3)}$, respectively. 
We may work to linear order in the probe field and write its wave equation as
\begin{equation}
\frac{\partial^2 E_1}{\partial x^2} + \frac{\partial^2 E_1}{\partial y^2} - \frac{1}{v^2}\left(1  +
 3 \epsilon_{2}\right) 
\frac{\partial^2 E_{1}}{\partial t^2}  = 0\,.
\label{eq:we1}
\end{equation}
Here $v$ is the effective speed of light in the medium when only linear effects take place (i.e., 
when $\chi^{(3)}=0$),
\begin{equation}
 v = \frac{c}{\sqrt{1 + \chi^{(1)}}} \, ,
 \label{speed}
 \end{equation}
and
\begin{equation}
\epsilon_{2} = \frac{\chi^{(3)}}{1 + \chi^{(1)}}\,E_0^2(t,x,y) \,.
\label{eq:eps2}
\end{equation}

The effect of the background field is to cause the probe field to experience an effective index of refraction
which depends upon space and time. This effect upon the propagation of probe
pulses is analogous to the effects of a gravitational field upon light propagation. Because the wavelength of the
probe field is short compared to the scale of variation of $\epsilon_{2}$, we may use a geometric optics
treatment with  a local refractive index defined as $n = c/v_{ph}$, where $v_{ph}$ is the 
velocity of the probe field and can be obtained directly from Eq.~(\ref{eq:we1}) as,
\begin{equation}
v_{ph} = \frac{v}{\sqrt{1  + 3\epsilon_2}}.
\label{eqvelocity1}
\end{equation}
If  $\epsilon_2 \ll1$, the refractive index can be expanded as 
\begin{equation}
n \approx n_0 \, \left(1 + \frac{3}{2} \epsilon_2 \right) \,,
\label{eq:n}
\end{equation}
where $n_0 = c/v = \sqrt{1 + \chi^{(1)}}$ is the refractive index of the medium when only linear effects are included. 

An important limitation of our model is the presence of dispersion in realistic materials, while gravity is dispersionless.
However, many materials including nonlinear ones have optical parameters which are relatively independent of
frequency over a broad range. This typically occurs at frequencies below that of any resonances, that is, in the infrared
part of the spectrum. 

So far our discussion has been at the classical level. Now we wish to regard the background field  as
a quantized field operator  ${\E}_0 (t,{ \x})$ , and to replace $E_0^2$ in the above expressions by an
expectation value of the normal ordered square of this operator:
\begin{equation}
E_0^2 \rightarrow \langle :{{\E}_0}^2 : \rangle \,.
\end{equation}
The effective refractive index becomes an expectation value of Eq.~\eqref{eq:n}:
\begin{equation}
\langle n \rangle = n_0\left(1+ \frac{3}{2}\, \langle\epsilon_2\rangle \right)
= n_0 +  \frac{3\chi^{(3)}}{2 n_0} \,\langle   :{{\E}_0}^2 :   \rangle \,.
\label{eq:n2}
\end{equation}
Note that we are not including any effects of vacuum fluctuations of the quantized electric field.
In the case that the field is in its vacuum state, we set $\langle n \rangle = n_0$. We may view
$n_0$ as already including any vacuum fluctuations corrections to the effective refractive index.

In quantum states, such as coherent states, for which $ \langle :{{\E}_0}^2 : \rangle > 0$ everywhere,
we find $\langle n \rangle > n_0$. This is the classical behavior in which the effect of the background field
is to slow the speed of probe pulses, and is analogous to the gravitational effects of positive energy
density. However, in nonclassical states, such as squeezed vacua, it is possible to have 
$ \langle :{{\E}_0}^2 : \rangle < 0$ in finite regions. In this case, $\langle n \rangle < n_0$, so
the effect is to increase the speed of pulses. 
This is analogous to the gravitational effects of exotic matter leading to superluminal propagation.

\section{Subvacuum Effects from Squeezed States}
\label{sec:squeeze}

\subsection{Mean Squared Electric Field in a Multimode Squeezed Vacuum}
\label{sec:multi}

Here we turn to the explicit construction of $ \langle :{{\E}_0}^2 : \rangle$ for the case of plane wave
modes in a squeezed vacuum state. The quantum field ${E}_0(t, \x)$  can be expanded  as 
\begin{equation}
 {\E}_0 (t,{ \x}) = \sum_{\k \lambda} \left[ a_{\k \lambda} \, \hat{\bf e}_{\k \lambda}\, g_{\k} (t,{\x}) + 
  a^\dagger_{\k \lambda}  \,   \hat{\bf e}^\ast_{\k \lambda}    g^*_{\k} (t,{\x}) \right]  \,.
\label{eq:Efield}
\end{equation}
Here $\hat{\bf e}_{\k \lambda}$ is  polarization vector and $ a^\dagger_{\k \lambda}$ and 
${a}_{\k \lambda}$ are photon creation and annihilation operators, obeying the commutation relation 
$[{a}_{\k \lambda}, a^\dagger_{\k' \lambda'}] = \delta_{\k,\k'} \,\delta_{\lambda,\lambda'} $.
We take the mode function to be
\begin{equation}
g_{\k}(t,\x) = \sqrt{\frac{\hbar\omega}{2\epsilon_0V}}\; {\rm e}^{i(\k\cdot\x -\omega t)}\, ,
\end{equation}
where $V$ is the quantization volume and $\omega = v /|\k|$. This function is  a solution of the Maxwell
equations for a medium with index of refraction $n_0$. 

We will assume that all of the excited modes are linearly polarized in the $z$ direction, so 
$\hat{\bf e}_{\k \lambda} = \hat{\bf z}$ for these modes. We further assume that the quantum state is
a multimode squeezed vacuum state, which may be constructed as
\begin{equation}
|\psi\rangle  = \prod_{\k}  S[\zeta_\k]  |0\rangle\,,
\label{eq:squeeze}
\end{equation}
with the product taken over all excited modes.
Here $S[\zeta_\k]$ represents the squeeze operator for mode $\k$, defined by
\begin{equation}
S[\zeta_\k] = \exp \left[\frac{1}{2}\left(\zeta_\k^* a_\k^2 -\zeta_\k a^{\dagger 2}_\k\right)\right]\,,
\label{eq:squeeze-op}
\end{equation}
where $\zeta_\k = q_\k {\rm e}^{i \eta_\k}$ denotes the complex squeeze parameter.
In this state, we find
\begin{equation}
\langle a_\k^{\dagger} {a_{\k'}}\rangle = \delta_{\k,\k'} \,\sinh^2 q_k \,,
\end{equation}
and
\begin{equation}
\langle  a_\k a_{\k'} \rangle =   \langle a_\k^\dagger a_{\k'}^\dagger \rangle^* =   
-  \delta_{\k,\k'} \,{\rm e}^{i\eta_\k}\cosh q_\k \sinh q_\k \,,
\end{equation}
where the polarization label $\lambda$ is now suppressed.

Using the above results, we find that the expectation value of the electric field is equal to zero,
$ \left\langle {\E}_0(t, \x)\right\rangle =0$. However, the expectation value of the squared
electric field operator becomes
\begin{equation}
\left\langle :\E_0^2(t,\x): \right\rangle =  \frac{\hbar}{\epsilon_0\,V}\,  \sum_{\k} \omega \,
\sinh q_{\k}\, \left[ \sinh q_{\k} +  \cosh q_{\k}\, \cos\left(2\omega t -2\k\cdot\x  -\eta_{\k}\right) \right] \,.
\label{eq:E2}
\end{equation}
In the limit of large quantization volume $V$, we may use
\begin{equation}
\frac{1}{V} \sum_{\k} \rightarrow  \frac{1}{(2\pi)^3} \int d^3 k \,.  
 \label{eq:bandwidth}
\end{equation}
We wish to consider the case where all of the excited modes are peaked in angular frequency about
$\omega = \Omega$ and in wavevector about $\k =(k,0,0)$ with a small but finite bandwidth. This describes a nearly
monochromatic beam of squeezed light propagating approximately in the $x$-direction. In this case,
we have
\begin{equation}
\left\langle :\E_0^2(t,\x): \right\rangle = 
\frac{\hbar}{4\pi^2\epsilon_0}\,k^2\, \Delta k\Delta\theta\,\Omega\, \sinh q
\big[\sinh q + \cosh q\cos(2\Omega t -2 \k\cdot\x- \eta)\big] \,,
\label{eq:field2}
\end{equation}
where $\Delta k$ denotes the bandwidth in wavenumber, and $\Delta\theta$ denotes the angular spread 
around the  $x$-direction, with $\Delta k /k \ll 1$ and $\Delta\theta\ll 1$. Here we have assumed that the
squeezed state parameters $q$ and $\eta$ are approximately constant within the bandwidth of the
excited modes.
If we define
\begin{equation}
\alpha = \frac{3\hbar}{16\pi^2\epsilon_0}\left[\frac{\chi^{(3)}}{1 + \chi^{(1)}}\right]\Omega k^2\Delta k\Delta\theta\, ,
\label{eq:alpha1}
\end{equation}
then the fractional change in the index of refraction becomes
\begin{equation}
\frac{\langle{n}\rangle - n_0}{n_0} = 
2\alpha \,\left[\sinh^2q + \sinh q\cosh q\cos(2\Omega t  - 2 \k\cdot\x- \eta)\right] \,.
\label{eq:devia1}
\end{equation}

\begin{figure}
 \centering
 \includegraphics[scale=1]{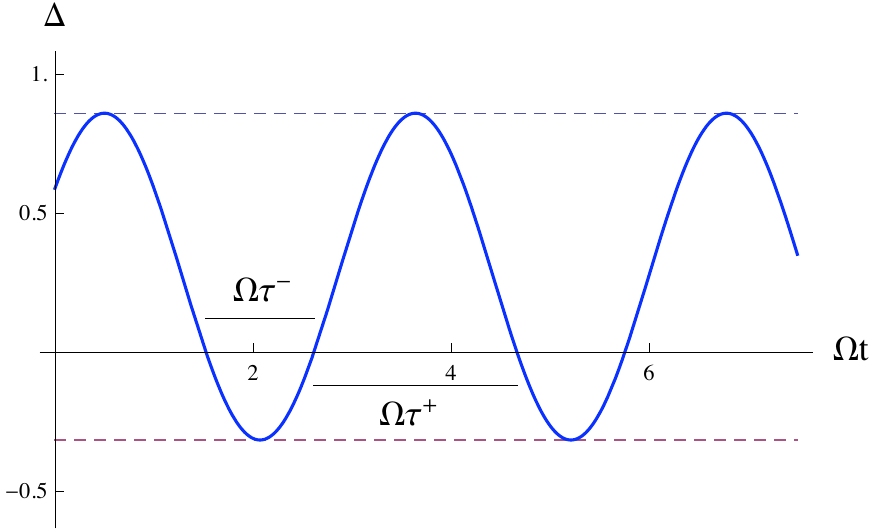}
 \caption{Behavior of the quantity $\Delta$ as function of time. Notice that subvacuum effects ($\Delta <0$) 
 occur periodically in time intervals of duration $\tau^{-}$. In this plot we have set $q=0.5$.}
 \label{fig1}
 \end{figure}

It is possible for this fractional change to be negative, a subvacuum effect. 
For small squeeze parameter $q \ll 1$, Eq.~(\ref{eq:devia1}) reads
\begin{equation}
\frac{\langle{n}\rangle - n_0}{n_0}  \approx 2\alpha \, q \,\cos(2\Omega t  - 2 \k\cdot\x -\eta)\,.
\end{equation}
In this limit, $\langle{n}\rangle - n_0 < 0$ for half of the time, but its magnitude is small, being
of first order in $q$. In the opposite limit of large squeeze parameter $q \gg 1$, we have
\begin{equation}
\frac{\langle n \rangle - n_0}{n_0} \approx 
\frac{1}{2} \alpha\, \left\{ {\rm e}^{2q} \,\left[1 - \cos(2\Omega t - 2 \k\cdot\x -\eta)\right]   -2 \right\} \,.
\end{equation}
In this limit $\langle{n}\rangle - n_0 < 0$ only for a short interval of time when 
$ \cos(2\Omega t - 2 \k\cdot\x -\eta) \approx 1$, but it can be arbitrarily negative for large $q$.
This inverse relation between the magnitude and duration of a subvacuum effect is an illustration
of the quantum inequalities.
The sign of $\langle{n}\rangle - n_0$ is determined by the sign of
\begin{equation}
\Delta = \sinh q[\sinh q+ \cosh q\cos(2\Omega t  - 2 \k\cdot\x- \eta)]
= \sinh q \, \cosh q[\tanh q+ \cos(2\Omega t  - 2 \k\cdot\x- \eta)] \,.
\end{equation}
Then $\Delta  <  0$ during a time interval of duration
\begin{equation}
\tau^{-} = \frac{\pi}{\Omega} - \frac{1}{\Omega}\arccos(-\tanh q)\,,
\end{equation}
and $\Delta  >  0$ during an interval of
\begin{equation}
\tau^{+} = \frac{1}{\Omega}\arccos(-\tanh q)\,.
\end{equation}
These intervals are illustrated in Fig.~\ref{fig1}.

We define $\langle  n\rangle^+$ as the largest value that $\langle{n}\rangle$ can take 
\begin{equation}
\langle  n\rangle^+ = n_0\left(1 + 2\alpha \,e^q\sinh q\right).
\label{eq:n+}
\end{equation}
Likewise, we define $\langle  n\rangle^-$ as its smallest value 
\begin{equation}
\langle  n\rangle^- =  n_0\left[1 - \alpha \,\left(1 - e^{-2q}\right)\right].
\label{eq:n-}
\end{equation}
These correspond to the maxima and minima of $\Delta$, respectively, which are illustrated in Fig.~\ref{fig1}.

\subsection{Refraction: A Magnitude Estimate}
\label{sec:refract}

\begin{figure}
 \centering
 \includegraphics[scale=0.4]{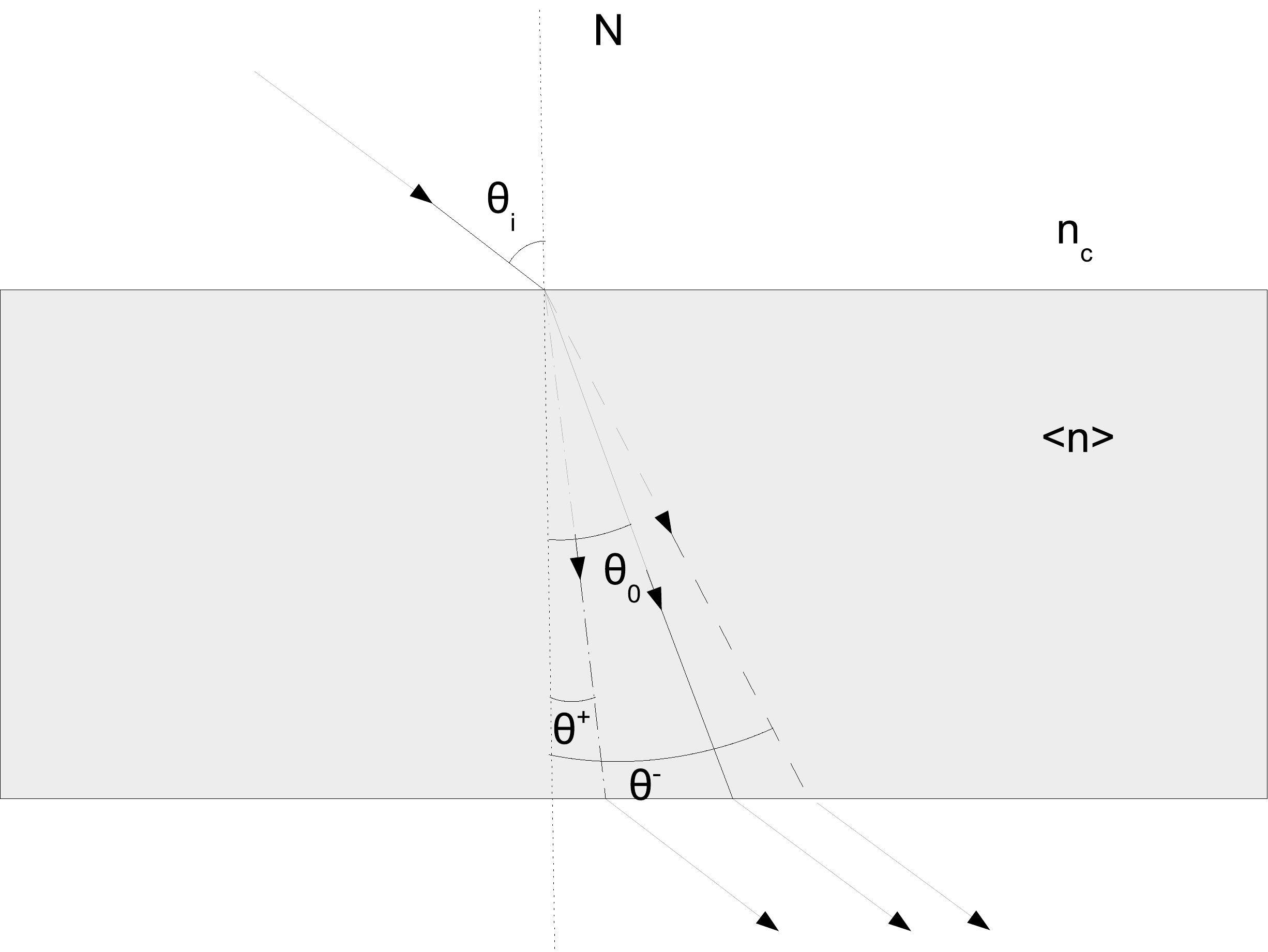}
 \caption{The refraction of the probe ray in the nonlinear medium is illustrated. In the absence of the background
 field, the angle of refraction is $\theta_0$. When the index of refraction attains its maximum value, the angle becomes
 $\theta^+$, and at its minimum value, the angle becomes $\theta^-$. The former case models the gravitational effect
 of positive energy, and the latter that of negative energy.}
 \label{fig2}
 \end{figure}

Now consider refraction at an interface. Suppose that a ray is incident at an angle of $\theta_i$ from a material
with index of refraction $n_c$ into the nonlinear medium with mean index of refraction $\langle  n\rangle$,
as shown in Fig.~\ref{fig2}. When $\langle n \rangle =  n_0$, the angle of refraction is $\theta_0$, given by
Snell's law to be
\begin{equation}
n_0\, \sin \theta_0 = n_c\, \sin \theta_i\,.
\end{equation}
In general, $\langle n(\x,t) \rangle$ is a function both of position and time. This will cause the trajectories of the
refracted rays in the nonlinear material to be curved.  However, the scale for this curvature is of the order of the
wavelength  associated with the background field, which is assumed long compared to the wavelength of the
probe field. Thus there is a scale longer than the probe field wavelength on which the refracted rays are 
approximately straight, as depicted in  Fig.~\ref{fig2}. Let $\theta^+$ be the angle of refraction when
$\langle  n\rangle$ takes its maximum value, $\langle  n\rangle \approx \langle  n\rangle^+$. Similarly, let
$\theta^-$ be the angle of refraction at the minimum value, 
$\langle  n\rangle \approx \langle  n\rangle^-$. These two cases correspond to the maximum and minimum
speeds of light in the material, respectively. In particular, $\theta^-$ is the angle of the ``superluminal"
ray which travels faster than the normal speed of light in the material.

Here we wish to find a numerical estimate for 
\begin{equation}
\delta\theta^- = \theta^- - \theta_0 \,,
\end{equation}
which is a dimensionless measure of the magnitude of the subvacuum effect. Snell's law gives
\begin{equation}
n_0\, \sin \theta_0 = n^-\, \sin \theta^- \,.
\end{equation}
If we expand to first order in $\alpha$, we have
\begin{equation}
\delta\theta^-  = \frac{n_c}{n_0^2} \, \sin \theta_i \, \frac{\delta n^-}{\cos \theta_0} \,,
\end{equation}
where 
\begin{equation}
\delta n^- = \langle n \rangle^- - n_0 = - n_0\, \alpha \,\left(1 - e^{-2q}\right) \,.
\end{equation}
This result may be expressed as
\begin{equation}
\delta \theta^- = \alpha \,\left(1 - e^{-2q}\right) \, \tan\theta_0 \,,
\end{equation}
or as
\begin{equation}
\delta n^- \approx \alpha  \tan\theta_0 \,,
\end{equation}
for $q \agt 1$.
The latter form reflects the state independent quantum inequality lower bound on $\langle n \rangle$.
 Squeezed states with a squeezing level
of $10\, {\rm db}$ have been created experimentally~\cite{sq2}. This corresponds to about $5$ photons 
per mode, or $\sinh q \approx 5$ and $q \approx 1.5$, leading to
\begin{equation}
1 - e^{-2q} \approx 0.95 \,.
\label{eq:10db}
\end{equation}
Because we can arrange to have $\tan \theta_0$ of order unity, the magnitude of $\delta n^-$ is determined 
by  $\alpha$. We may rewrite Eq.~\eqref{eq:alpha1} as
\begin{equation}
\alpha = 1.06 \times 10^{-7} \, n_0 \, \left(\frac{\chi^{(3)}}{10^{-18} {\rm m^2/V^2}} \right) \,
\left(\frac{1\,\mu{\rm m}}{\lambda} \right)^4 \, \frac{\Delta k}{k} \, \Delta\theta \,, 
\label{eq:estimate}
\end{equation}
where $\lambda = c/\Omega$ is the vacuum wavelength of background field. Third order susceptibilities
with $\chi^{(3)} \agt 10^{-18} {\rm m^2/V^2}$ in the infrared region can be obtained in materials such
as silicon and germanium~\cite{Hon11}. Furthermore, $\chi^{(3)}$ for these materials is relatively independent
of wavelength for $\lambda \agt 4\,\mu{\rm m}$, so the nonlinearity becomes approximately dispersionless.
Although the fractional changes in speed or deflection angle are small, it is conceivable that they could be
observed.

\subsection{Lower Bounds on the Time Averaged Mean Squared Electric Field}
\label{sec:bounds}

It is natural to ask whether the multimode squeezed vacuum state used in writing Eq.~\eqref{eq:field2} gives the best
possible decrease in refractive index, or if there exist quantum states which can do much better. Here we address the
related question of the lower bound on $\langle E^2 \rangle$ averaged with a Lorentzian sampling function in time.
Let $g(t,\tau)$ be a Lorentzian in $t$ with width $\tau$:
\begin{equation}
g(t,\tau) = \frac{\tau}{\pi(t^2+\tau^2)} \,,
\end{equation}
so $\int_{-\infty}^\infty g(t,\tau) \,dt =1$ for all $\tau$. The Lorentzian average of Eq.~\eqref{eq:field2} at $\x =0$
becomes
\begin{equation}
\int_{-\infty}^\infty g(t,\tau)\, \langle :\E_0^2(t,0): \rangle \,dt =  
\frac{\hbar}{4\pi^2\epsilon_0}\,k^2\, \Delta k\Delta\theta\,\Omega\, \sinh q\, \cosh q\,
(\tanh q + \cos \eta\, {\rm e}^{-2\Omega \tau} )\,.
\label{eq:LorAv}
\end{equation}
It is clear that the right hand side of this expression attains its minimum value when $\cos \eta =-1$, so we can
write
\begin{equation}
\int_{-\infty}^\infty g(t,\tau)\, \langle :\E_0^2(t,0): \rangle \,dt \geq 
\frac{\hbar}{4\pi^2 \epsilon_0 c^3 \, \tau^4}\, \frac{\Delta u}{u}\,\Delta\theta\, f(q,u) \,,
\end{equation}
where we have set $k =\Omega/c$ for empty space, and define $u= \Omega \tau$ and
\begin{equation}
f(q,u) = u^4\, \sinh q \, \cosh q\, (\tanh q - {\rm e}^{-2 u}) \,.
\end{equation}
Plots of the function $f(q,u)$ reveal that it attains its minimum value of $f \approx -0.0047$ at $u \approx 1.0$
and $q \approx 0.07$. This leads to a lower bound on the Lorentzian average in our choice of squeezed state
of
\begin{equation}
\int_{-\infty}^\infty g(t,\tau)\, \langle :\E_0^2(t,0): \rangle \,dt \geq B =
 -\frac{1.2 \times 10^{-4}\, \hbar}{ \epsilon_0 c^3 \, \tau^4} \, \frac{\Delta u}{u}\,\Delta\theta\,. 
 \label{eq:lower}
\end{equation}
Compare this with the estimated optimum bound for all quantum states, obtained in Eq.~(73) of Ref.~\cite{FFR12},
which may be expressed as
\begin{equation}
B_{\rm opt} =
 -\frac{3.0 \times 10^{-4}\, \hbar}{ \epsilon_0 c^3 \, \tau^4} \,. 
\end{equation}
Although we previously assumed that $\Delta\theta \ll 1$ and ${\Delta k}/{k} = {\Delta u}/{u} \ll 1$, for the 
present purpose, we could relax these restrictions.
If we allow integration over a sufficiently wide range of frequencies and directions, then we might have 
$\Delta\theta \approx \pi$ and ${\Delta u}/{u}$ of order unity, so $B \approx B_{\rm opt}$.  In this case,
the Lorentzian average in our choice of squeezed state is close to the allowed bound for any state. 
Note that the value of $q \approx 0.07$ which leads to Eq.~\eqref{eq:lower} is smaller than the value used
in writing Eq.~\eqref{eq:10db}. This implies that extremizing a Lorentzian average and extremizing 
$\delta \theta^-$ are not quite the same. Nonetheless, that fact that the multimode squeezed state does
a good job with the former may suggest that 
one cannot significantly improve the estimate in Eq.~\eqref{eq:estimate} by changing the quantum state.

\section{Casimir Type Effects}
\label{sec:Casimir}

In this section, we wish briefly to discuss the possibility of effects which arise from boundary conditions
imposed upon the quantized electromagnetic field. The simplest example of this type of effect is that of
a single plane mirror with perfect reflectivity. In this case, the mean squared electric field is positive:
\begin{equation}
\langle \E^2 \rangle = \frac{3 \, \hbar c}{16\pi^2\epsilon_0\, z^4} \,,
\label{eq:E2-plate}
\end{equation}
where $z$ is the distance to the mirror, and $\langle \E^2 \rangle$ is understood as the shift relative to the
empty space vacuum state.  However, the  mean squared magnetic field is negative:
\begin{equation}
\langle \B^2 \rangle = -\frac{3 \, \hbar }{16\pi^2 c \epsilon_0\, z^4} \,.
\end{equation}
Despite the presence of the mirror, the squares of the individual field components exhibit isotropy
(These results may be obtained from formulas in Sect.~3 of Ref.~\cite{BM69}.):
\begin{equation}
\langle E_x^2 \rangle = \langle E_y^2 \rangle = \langle E_z^2 \rangle = \frac{1}{3}\, \langle \E^2 \rangle\,,
\label{eq:Ex2}
\end{equation}
and
\begin{equation}
\langle B_x^2 \rangle = \langle B_y^2 \rangle = \langle B_z^2 \rangle = \frac{1}{3}\, \langle \B^2 \rangle\,.
\end{equation}
(This unexpected result is analogous to the spatial isotropy of the Schwarzschild geometry, which is evident
when the  Schwarzschild metric is written in isotropic coordinates.) For metal mirrors, the above forms are good
approximations when $z$ is large compared to the plasma wavelength, but are modified at shorter
distances~\cite{SF02,SF05}. 

In order to produce an increase in the speed of light due to third order nonlinear effects near this mirror, 
one would need a
material with a nonzero magnetic  $\chi^{(3)}$. However, materials with a nonzero electric  $\chi^{(3)}$
will exhibit a reduction in light propagation speed near a mirror. This is analogous to the gravitational effect
of a vacuum energy which satisfies the null energy condition. We can estimate the magnitude of this reduction
for modes polarized in the $x$-direction by combining Eq.~\eqref{eq:n2}   with  Eqs.~\eqref{eq:E2-plate} and 
\eqref{eq:Ex2} to write the fractional change in refractive index as
\begin{equation}
\frac{\langle{n}\rangle - n_0}{n_0} = \frac{3\chi^{(3)}}{2 n_0} \,\langle   E_x^2   \rangle 
= \frac{3 \hbar c \,\chi^{(3)}}{32 \pi^2 \epsilon_0 n_0^2\, z^4}
= \frac{3.4 \times 10^{-11}}{n_0^2}\,  \left(\frac{\chi^{(3)}}{10^{-18} {\rm m^2/V^2}} \right) \,
\left(\frac{1\,\mu{\rm m}}{z} \right)^4 \,.
\end{equation}
This reduction in light speed is an analog of the Shapiro time delay~\cite{Shapiro},  but produced by 
quantum vacuum effects. Simple geometries such as the plane mirror seem to result in
$\langle \E^2 \rangle > 0$. However, it would be of interest to find geometries where 
$\langle \E^2 \rangle < 0$, leading to an increase in light speed.

\section{Summary and Discussion}
\label{sec:sum}

We have presented an analog model, using nonlinear optics, for the gravitational effects of the expectation value
of a quantum stress tensor upon light propagation. The key ingredient in our model is a material with
nonzero third order polarizability and with a nonzero expectation value of the squared electric field operator.
This expectation value in our model plays the role of the expectation value of the quantum stress tensor
in semiclassical gravity. The origin of $\langle E^2 \rangle$ is a background field in a squeezed vacuum
state, or a Casimir vacuum state. This quantity alters the propagation speed of a probe pulse. When
$\langle E^2 \rangle > 0$, the speed is reduced, in analogy to the gravitational effects of ordinary matter.
When $\langle E^2 \rangle < 0$, the speed is increased, in analogy to the gravitational effects of exotic matter  
which violates the null energy condition. The estimates of the magnitude of the effect given in 
Sect.~\ref{sec:squeeze} indicate that changes in speed or deflection angle as large as $10^{-7}$
might be achievable in realistic experiments and hence possibly detectable. Such a detection would
be of interest in its own right as an effect in quantum optics, as well as a way to study the otherwise
very small quantum effects in gravity.

In this paper, we have only been concerned with the expectation value,   $\langle E^2 \rangle$. However, the
squared electric field operator undergoes subtle fluctuations both in the vacuum~\cite{FFR12} and in other 
quantum states. In a nonlinear material, these will lead to fluctuations of the path of light rays which will
model passive quantum fluctuations of gravity, those due to quantum stress tensor fluctuations. These 
effects will be studied in a future paper.

 \begin{acknowledgments}
This work was supported in part by the National Science Foundation under Grant PHY-1205764,
and by the Brazilian research agencies
CNPq, FAPEMIG and CAPES under scholarships BEX 18011/12-8 and CNPq 245985/2012-3.
\end{acknowledgments}

\end{document}